\documentclass{iau}

\usepackage{amsmath}
\usepackage{graphicx}
\usepackage{multirow}

\newcommand{\teff}{$T_{\mathrm{eff}}$}

\newcommand{\logg}{\mbox{log \textit{g}}}

\begin{document}

\lefttitle{Cambridge Author}
\righttitle{Proceedings of the International Astronomical Union: \LaTeX\ Guidelines for~authors}

\jnlPage{1}{7}
\jnlDoiYr{2025}
\doival{10.1017/xxxxx}
\volno{395}
\pubYr{2025}
\journaltitle{Stellar populations in the Milky Way and beyond}

\aopheadtitle{Proceedings of the IAU Symposium}
\editors{J. Mel\'endez,  C. Chiappini, R. Schiavon \& M. Trevisan, eds.}

\title{How homogeneous was the chemical enrichment of the Milky Way 13 gigayears ago?}

\author{Riano E. Giribaldi$^1$, Laura Magrini$^1$, Martina Rossi$^{2,3}$, Anish Amarsi$^4$, Davide Massari$^3$, Donatela Romano$^3$}
\affiliation{$^1$INAF – Osservatorio Astrofisico di Arcetri, Largo E. Fermi 5, 50125
Firenze, Italy \email{riano.escategiribaldi@inaf.it}\\
$^2$Dipartimento di Fisica e Astronomia, Alma Mater Studiorum, Università di Bologna, Via Gobetti 93/2, 40129 Bologna, Italy\\
$^3$INAF, Osservatorio di Astrofisica e Scienza dello Spazio, Via Gobetti 93/3, 40129 Bologna, Italy\\
$^4$Theoretical Astrophysics, Department of Physics and Astronomy, Uppsala University, Box 516~SE-751~20 Uppsala, Sweden}

\begin{abstract}
We reanalyze the chemical composition of the metal-poorest tail of the Galactic halo using highly accurate atmospheric parameters \citep{giribaldi2021, giribaldi2023} and cutting-edge 3D NLTE models \citep{amarsi2018}. Most [Mg/Fe] versus [Fe/H] diagrams in the literature exhibit significant scatter at [Fe/H] $\lesssim -2$~dex, often interpreted as evidence of inhomogeneous enrichment during the early phases of galaxy evolution \citep[e.g.,][]{rossi2021}. However, our analysis of observational data reveals that in the range $-3.5 <$ [Fe/H] $< -2$~dex, the [Mg/Fe] versus [Fe/H] distribution is relatively narrow. This finding suggests a low degree of stochastic enrichment in magnesium during these epochs in the Milky Way halo.
\end{abstract}


\maketitle

\section{Introduction}

The Milky Way is thought to have formed through multiple merger events over timeincluding at least one major accretion event approximately 9.5 Gyr ago involving the Gaia-Enceladus-Sausage (GES) progenitor \citep[e.g.,][]{belokurov2018, helmi2018, giribaldi2023a}.
The chemical signature of GES is evident in several abundance planes, such as [Mg/Fe] versus [Fe/H], [Ni/Fe] versus [(C+N)/O], and [Mg/Mn] versus  [Al/Fe] \citep[e.g.][]{feuillet21,montalban2021,horta2021,giribaldi2023}, largely thanks to the use of data from large spectroscopic surveys (e.g. GALAH and APOGEE).
However, these surveys observed few metal-poor stars ([Fe/H] $< -2$~dex), which are crucial  for probing the Galaxy's earliest evolutionary phases.
Until recently, most chemical analyses of metal-poor stars relied on  one-dimensional Local Thermodynamic Equilibrium (1D LTE) models applied to moderate signal-to-noise (S/N) spectra of red giant stars. 
Numerous studies in the literature report significant scatter in the [Mg/Fe] versus [Fe/H] plane, typically attributed to stochastic enrichment \citep[e.g.,][]{rossi2021}.
Nevertheless,  analyses of dwarf stars \citep{arnone2005} and the adoption of Non-Local Thermodynamic Equilibrium (NLTE) models for giant stars \citep{and2010} suggest that this scatter could partially arise from oversimplified assumptions in spectral analysis, introducing biases.
Here we show the results of our analysis of high-quality spectra of metal-poor dwarf and giant stars  employing  1D LTE, 1D NLTE, 3D LTE and 3D NLTE synthesis. We demonstrate how the placement of these stars in the [Mg/Fe] versus [Fe/H] plane changes depending on the adopted model assumptions.

\section{Data and sample}
\subsection{Observational data and stellar parameters}

We utilized ESO UVES archival spectra of 15 stars in the TITANS I sample \citep[][dwarf stars]{giribaldi2021} and 13 stars in the TITANS II benchmark sample \citep[][red giants]{giribaldi2023}. These spectra have a  resolving power R $> 40 000$ and a S/N $> 100$, enabling us to derive [Mg/Fe] with a  precision between 0.05 and 0.10~dex.       
Additionally,  we re-derived the stellar parameters and [Mg/Fe] of 22 stars in the {\it Gaia}-ESO survey \citep{randich2022, gilmore2022}. The left panel in Fig.~\ref{figure1} displays the Kiel diagram of the analysed stars. 
We recall that our [Mg/Fe] scale is anchored to  accurate effective temperatures (\teff) from 3D NLTE H$\alpha$ profiles 
\citep{amarsi2018,giribaldi2021,giribaldi2023}. 
We derived [Fe/H] using Turbospectrum \citep{gerber2023} under NLTE assumptions.  
We confirmed  that the ionization equilibrium is satisfied when \logg\ from Mg triplet lines is used as input \citep{giribaldi2023}.

\subsection{Space distribution}
\begin{figure*}
    \includegraphics[scale=.5]{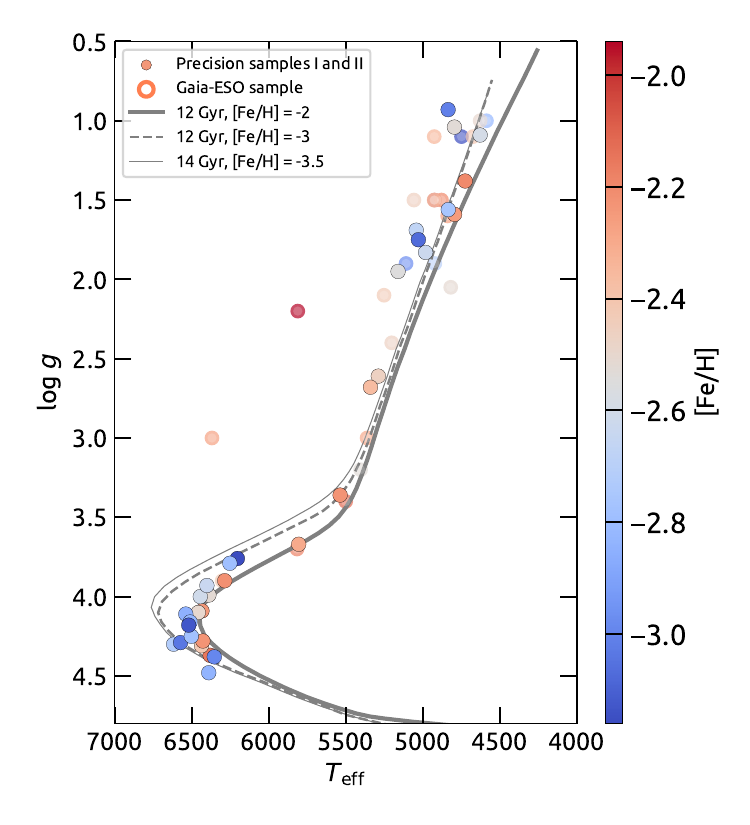}
    \includegraphics[scale=.45]{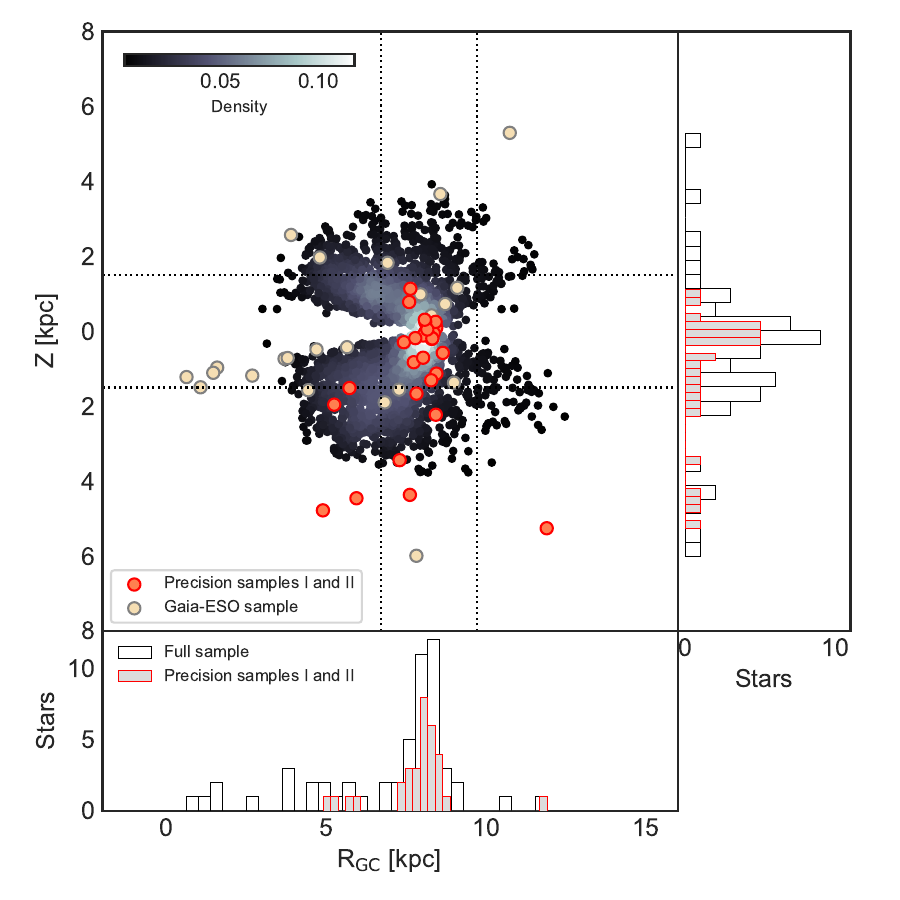}
    \caption{{\it Left panel:} Kiel diagram. Precision samples I and II, and the {\it Gaia}-ESO sample are displayed as indicated in the legends. Symbols are colour-coded according to the metallicity. Yonsey-Yale isochrones with ages and metallicity values given the legends are over-plotted as reference.
    {\it Right panel:} Spatial distribution of our sample stars with respect to the Galaxy centre. Stars in the GALAH survey are displayed as background reference.}
    \label{figure1}
  \end{figure*}

The right panel in Fig.~\ref{figure1} shows the location of our stars in the Galactocentric distance ($R_{GC}$) versus height  to the Galactic plane
($Z$) plane. Approximately 72\% (23 of 32) of the stars in the Precision
samples are located  within  the solar neighbourhood,
defined as a squared box of 3 kpc centred on the Sun. Among the {\it Gaia}-ESO sample, five of 22 stars are in the solar neighborhood, while 10 are positioned toward the Galactic center.

\subsection{Kinematic properties}

\begin{figure}
    \includegraphics[width=0.95\linewidth]{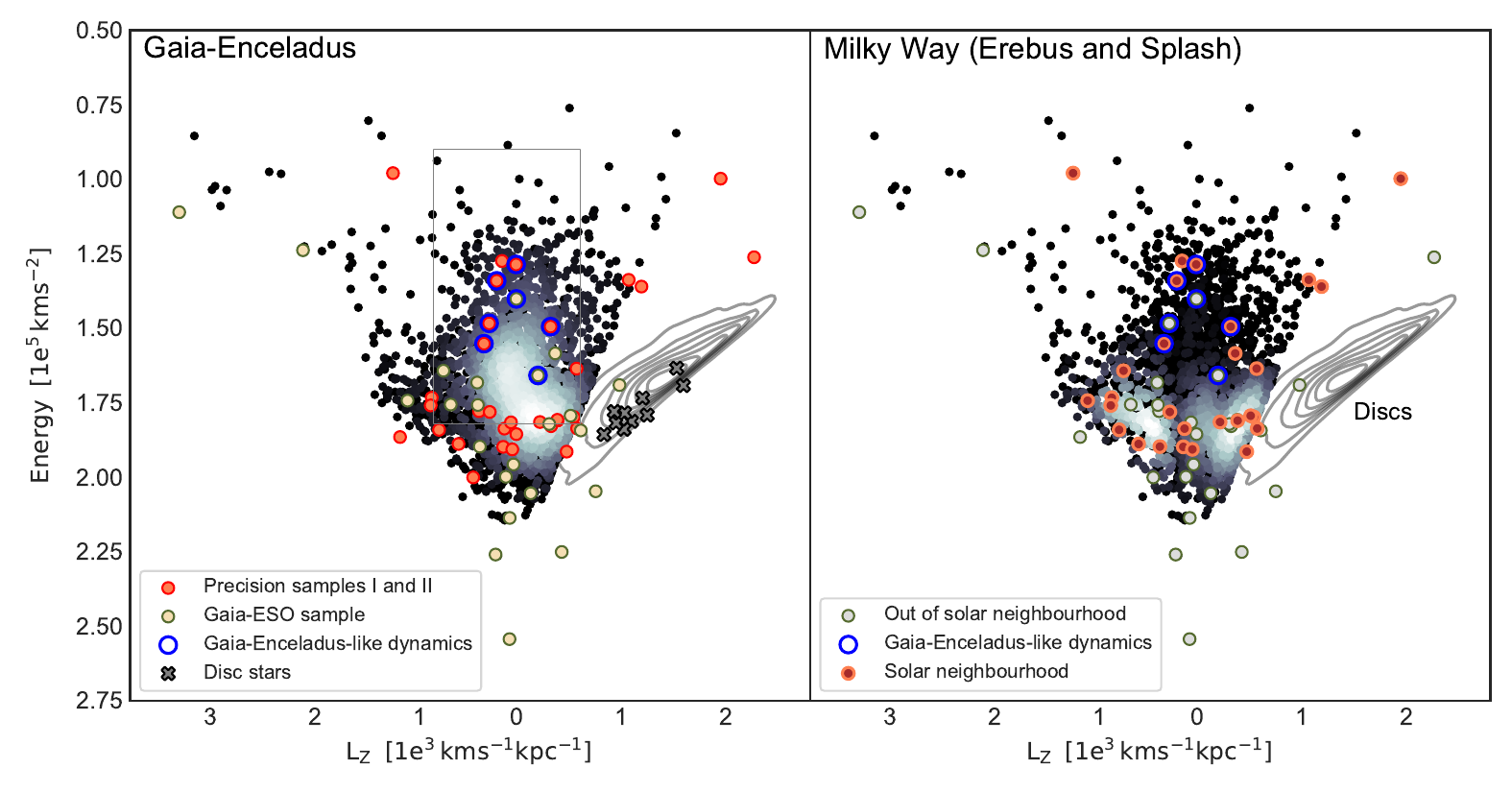}
    \caption{Lindblad diagram. Each panel colour-map emphasises the Gaia-Enceladus and Milky Way (Erebus and Splash) populations with the sample stars separated
    by \cite{giribaldi2023a}. {\it Left panel: } The symbols indicate to which sample the stars belong. Also, stars with Gaia-Enceladus-like dynamics are indicated by blue contours.
    The square box indicate the area with highest probability of enclosing GES stars defined by \cite{massari2019}.
    {\it Right panel: }The symbols indicate stars inside and outside the solar neighbourhood, as defined by the box in Fig.~\ref{figure1}. Stars with GES-like dynamics are indicated by blue contours as well.
        }
    \label{figure2}
  \end{figure}

Figure~\ref{figure2} shows our sample in the Lindblad diagram, with the GES and Milky Way populations (Erebus and Splash) identified by \cite{giribaldi2023a} displayed as a reference background. The distribution of our sample stars indicates they do not belong to any specific halo population but are instead widely dispersed. 

The left panel demonstrates that this broad distribution applies to both the precision samples and the \textit{Gaia}-ESO sample. The right panel further reveals that even stars located in the solar neighborhood exhibit a wide spread in the \(L_Z\) and bounding energy plane.

\section{Magnesium abundance under 1D LTE, 1D NLTE, 3D LTE, and 3D NLTE}

\begin{figure}
  \centering
    \includegraphics[width=0.48\linewidth]{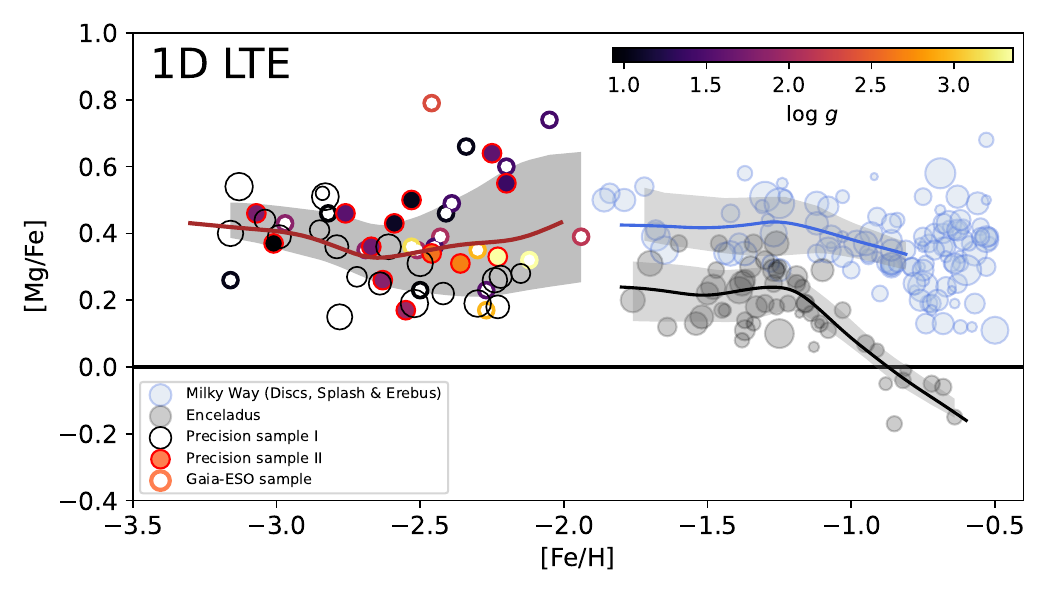}
    \includegraphics[width=0.48\linewidth]{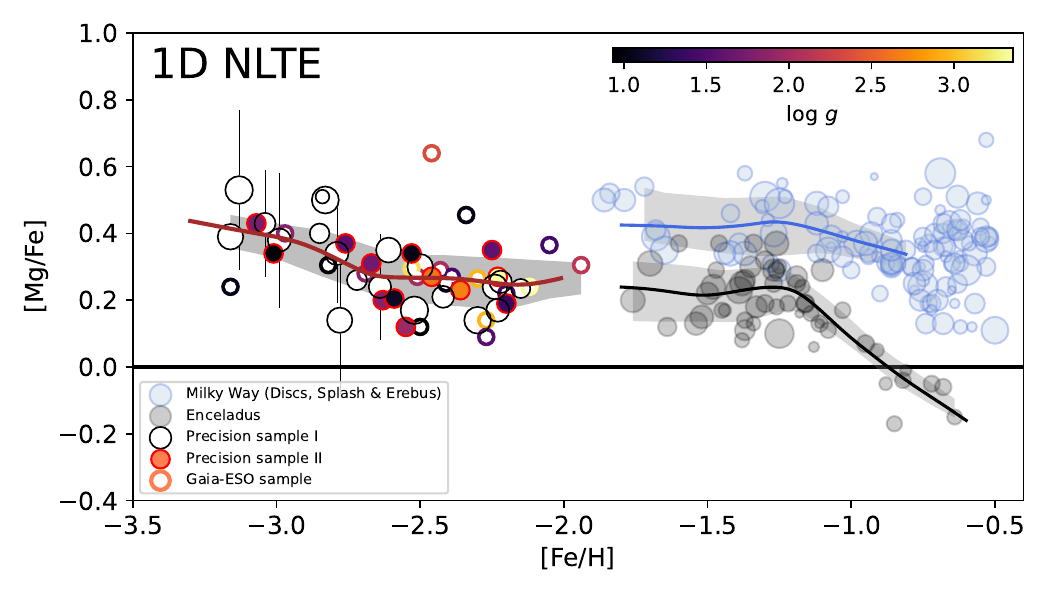}\\
    \includegraphics[width=0.48\linewidth]{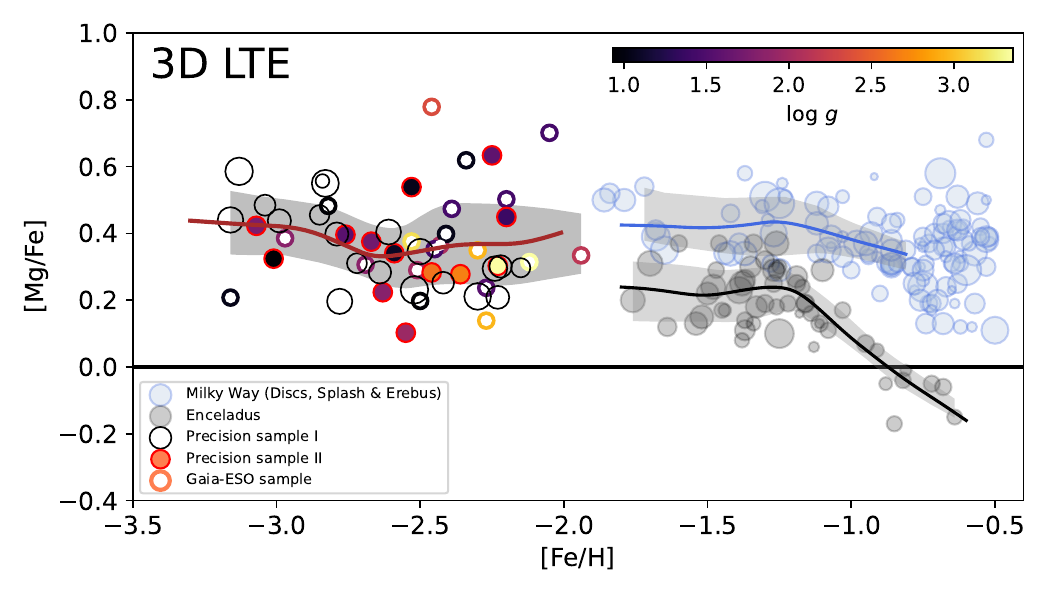}
    \includegraphics[width=0.48\linewidth]{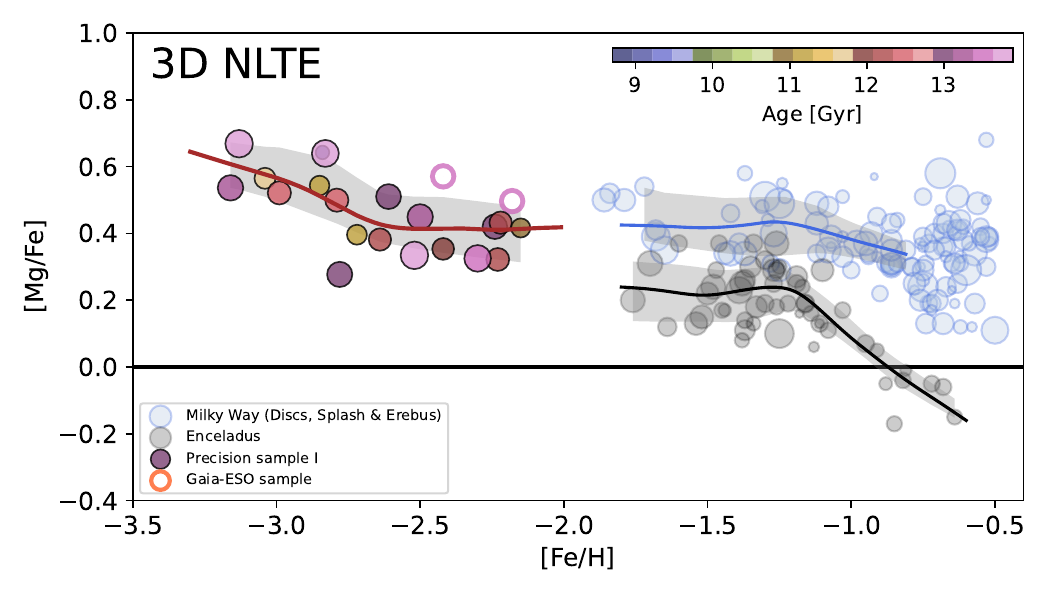}
    \caption{Symbols of dwarfs are size-coded according to their ages. As a reference, stars classified as in-situ Milky Way (blue) and Gaia-Enceladus (gray) in \cite{giribaldi2023a} are plotted. Red, blue, and black lines display LOWESS regressions. The shades display 20 to 80\% quantiles. 
    Dwarfs (Precision sample I) have no colour coding. Red giants (Precision sample II and Gaia-ESO sample) are colour-codded according to \logg. Their symbols follow the legends in the plots. 
        }
    \label{figure3}
  \end{figure}

We derived Mg abundance from the line at 5528~\AA\ and the {\it Gaia}-ESO linelist \citep{heiter2021}.
We performed line synthesis under 1D LTE and 1D NLTE using Turbospectrum and the departure coefficients based on the model atom in \cite{bergemann2017}.
We computed 3D LTE corrections using the \texttt{Scate} code \citep{hayek2011}, and 3D NLTE corrections (only for dwarfs) using the code \texttt{Balder} \citep{amarsi2018}.
Figure~\ref{figure3} shows the [Mg/Fe] - [Fe/H] diagrams obtained under 1D LTE, 1D NLTE, 3D LTE, and 3D NLTE assumptions.

\section{Summary and conclusions}

We confirm the hypothesis first proposed by \cite{arnone2005} that  most of the [Mg/Fe] dispersion typically observed in the literature arises from 1D LTE abundance determinations in giants with \logg\ $< 1.5$~dex. 
This is illustrated  in the top panels of Fig.~\ref{figure3}.  Notably, 3D LTE models do not reduce the [Mg/Fe] dispersion, as shown by comparing the top and bottom left panels in Fig.~\ref{figure3}.
In contrast, 1D NLTE models produce a narrow sequence of low-metallicity stars  in the [Mg/Fe]-[Fe/H] plane, which might represent  an extension of the GES  population (gray symbols and black line in Fig.~\ref{figure3}).
In addition, our dwarf stars are older than 12.5 Gyr (see bottom right panel in Fig.~\ref{figure3}) enabling us to trace magnesium enrichment during the earliest phases of the Milky Way's history, specifically within the first two Gyr after the Big Bang.
Abundances derived using 3D NLTE are shifted relative to those obtained with 1D NLTE, resulting in an offset of \(+0.2\)~dex  in the  [Mg/Fe] sequence. Consequently, the [Mg/Fe] level of the metal-poor star sequence becomes consistent with that of the Milky Way stars (blue symbols and blue line in Fig.~\ref{figure3}).

With our precise spectral analysis, we demonstrated that the [Mg/Fe] scatter at $-3 <$ [Fe/H] $<-2$~dex is remarkably narrow ($\sim$0.06~dex), suggesting that the dispersion reported in previous studies may have been artificially amplified by the LTE modeling of Mg lines. 
Using 3D NLTE models, the derived [Mg/Fe] ratios change significantly. For [Fe/H]~$> -2.8$dex, [Mg/Fe] abundances form a plateau around [Mg/Fe]$\approx 0.45$~dex, aligning the abundances of the metal-poorest populations with those of in-situ formed populations, such as Erebus \citep{giribaldi2023a} and Splash \citep[e.g.][]{belokurov2020,feuillet21}.

Below [Fe/H]~$< -2.8$~dex, a knee-like feature emerges, where [Mg/Fe] increases to approximately $0.65$dex at [Fe/H]$\sim -3.2$dex. This chemical pattern could arise from a single population, such as a merger with a low-mass galaxy, or from several populations with similar characteristics, corresponding to several merging building block galaxies with similar chemical properties. 

However, the kinematic properties of the current sample suggest a lack of coherence (Fig.\ref{figure2}). We propose that these stars may plausibly originate from a single population, whose spatial and kinematic properties were subsequently disrupted by multiple merger events.

\end{document}